\documentclass[twocolumn,amsmath,amssymb,superscriptaddress]{revtex4}

\usepackage[latin1]{inputenc}
\usepackage{upgreek,graphicx}
\usepackage{amsmath}
\usepackage{times}

\begin{document}

\title{Observation of anisotropic diffusion of light in compacted granular porous materials}

\author{Erik Alerstam}
\email{erik.alerstam@fysik.lth.se}
\affiliation{Department of Physics, Lund University, P.O. Box 118, 221 00 Lund, Sweden\\}
\author{Tomas Svensson}
\email{svensson@lens.unifi.it}
\affiliation{Department of Physics, Lund University, P.O. Box 118, 221 00 Lund, Sweden\\}
\affiliation{European Laboratory of Non-linear Spectroscopy (LENS), 50019 Sesto Fiorentino (Florence), Italy}

\begin{abstract}
It is known that compaction of granular matter can lead to anisotropic mechanical properties. Recent work has confirmed the link to pore space anisotropy, but the relation between compression, mechanical properties and material microstructure remains poorly understood and new diagnostic tools are needed. By studying the temporal and spatial characteristics of short optical pulses diffusively transmitted through compacted granular materials, we show that powder compaction can also give rise to strongly anisotropic diffusion of light. Investigating technologically important materials such as microcrystalline cellulose, lactose and calcium phosphate, we report increasing optical anisotropy with compaction force and radial diffusion constants being up to 1.7 times the longitudinal. This open new and attractive routes to material characterization and investigation of compression-induced structural anisotropy. In addition, by revealing inadequacy of isotropic diffusion models, our observations also have important implications for quantitative spectroscopy of powder compacts (e.g., pharmaceutical tablets). 
\end{abstract}

\maketitle

The physics of granular media in general, and compression and deformation in particular, is utterly complex \cite{Train1956_JPharmPharm,Goldenberg2005_Nature,Luding2005_Nature,Sun2011_JAdhesionSciTech}. Material anisotropy due to compression is one phenomena in this context that
remains poorly understood. It is well known that uniaxial compression of certain granular media can induce anisotropy in mechanical properties \cite{Ando1983_ChemPharmBull,Mullarney2006_IntJPharm}, and scanning electron microscopy, pulsed-gradient stimulated-echo NMR and spin-echo small-angle neutron scattering have recently confirmed that the pore structure itself indeed can be anisotropic \cite{Wu2008_PharmSciTech,Busignies2008_EJPB,Andersson2009_PowderTechnol,Porion2010_PharmRes}. The fundamental understanding of the interplay between compression, microstructure and anisotropy is, however, still in its cradle \cite{Andersson2009_PowderTechnol}.

In this work, we report that compression of granular matter also can give rise to anisotropic diffusion of light. Since the phenomenon is linked to pore space anisotropy, light scattering and diffuse spectroscopy may thus turn out to be a valuable, and non-destructive, tool for material characterization and fundamental investigations of compression-induced anisotropy. As the materials in which we observe anisotropic diffusion of light are of major technological importance, our findings also reveal an urgent need for better understanding of the microstructure and optics of compacted granular media. In particular, our results have important implications for quantitative spectroscopy of, e.g., compacted powders and pharmaceutical tablets. Anisotropy can, for example, be a  complication for spectroscopic methods that aim at separating the effects of scattering and absorption, methods which typically are based on isotropic diffusion models \cite{Shi2010a_JPharmSci}.

\begin{figure}[htb]
  \includegraphics[]{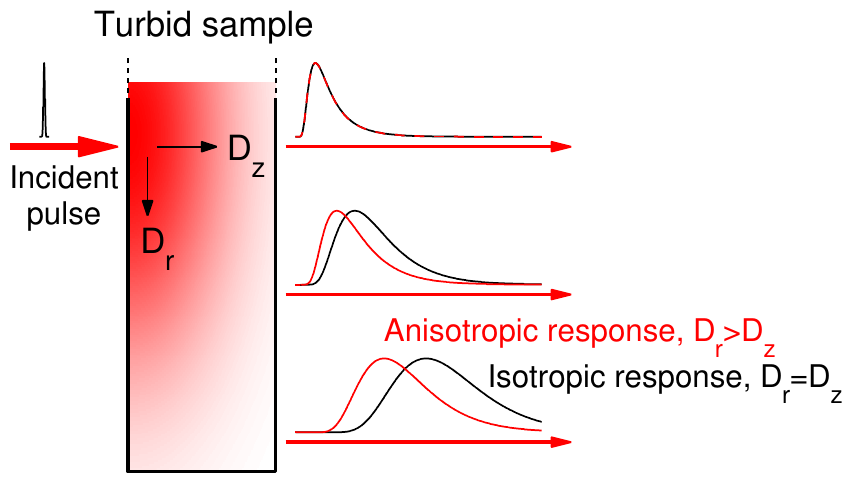}\\
  \caption{\textbf{Measurement principle.} In spatially resolved photon time-of-flight spectroscopy, anisotropic diffusion manifests itself as an anomalous spatial dependence of the arrival and shape of diffusively transmitted pulses. When radial diffusion is faster than longitudinal ($D_r>D_z$, in terms of diffusion constants), as for the cylindrical compacts involved in this study, pulses detected at non-zero radial positions arrive earlier and are narrower than what could be expected from looking only at on-axis transmission. In fact, the temporal shape of the pulse transmitted along the axis of incidence is independent of the radial diffusion constant. }\label{fig:concept}
\end{figure}

Anisotropic diffusion of light has previous been observed and studied in materials with rather evident anisotropic microstructure, including etched porous semiconductors \cite{Johnson2002_PRL}, nematic liquid crystals \cite{vanTiggelen1996_PRL,Wiersma1999_PRL,Wiersma2000_PRE,vanTiggelen2000_RevModPhys}, fibrous and stretched plastics \cite{Johnson2008_OptExpress,Johnson2009_JBO}, wood \cite{Kienle2008_OptExpress} and biological tissues such as muscle, teeth, bone, brain and skin \cite{Kienle2004_OptLett,Kienle2006_PhysRevLett,Binzoni2006b_PhysMedBiol}. In contrast, we report on anisotropic diffusion in materials where presence of microstructural anisotropy is far from self-evident, still being the subject to active research, and where light propagation conventionally is assumed to be isotropic. In fact, the radial symmetry of the diffusion occurring in the compacted granular systems discussed here makes it impossible to detect optical anisotropy by, e.g., employing steady-state methods and searching for spatial asymmetries in reflected or transmitted intensity profiles. To reveal the anisotropic diffusion of light, we have instead relied on spatially resolved measurements of light transport dynamics. The main concept behind our approach is illustrated in Fig. \ref{fig:concept}.

\section*{RESULTS}
The granular materials investigated are cylindrical compacts made by uniaxial compression of powders based on microcrystalline cellulose (MCC), lactose or calcium phosphate. All samples have a diameter of 13 mm and a total weight of 500 mg. The compaction force applied varied between 5 and 55 kN, and thickness were between 1.9 and 3.5 mm. The MCC-samples represent realistic pharmaceutical wet granulated compacts, and were manufactured from three different granule size fractions A, B and C (granules being $<150$ $\upmu$m, $150-400$ $\upmu$m and $>400$ $\upmu$m, respectively). Lactose and calcium phosphate samples were, on the other hand, made from pure powders (5-10 $\upmu$m mean particle size). For reference, we also make experiments on homogenous isotropic turbid materials with scattering properties similar to those of the granular samples. These are Spectralon (a commercially available porous fluoropolymer), a macroporous sintered alumina ceramic \cite{Svensson2010_OptLett,Svensson2011_PRL}, and a TiO$_2$-based epoxy phantom \cite{Svensson2009_RevSciInstrum}.

\begin{figure}[htb]
  \includegraphics[]{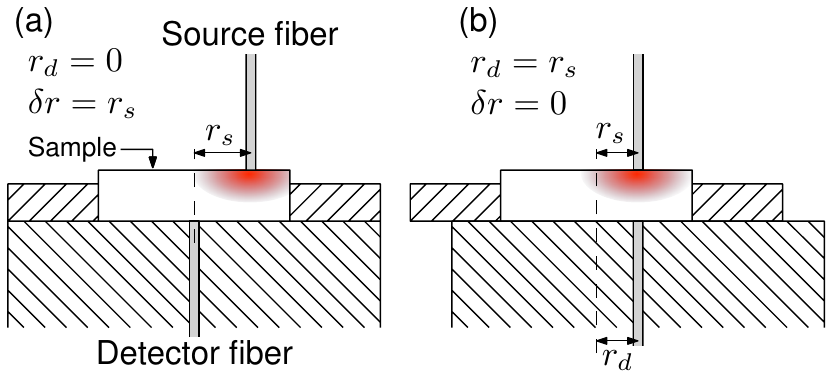}\\
  \caption{\textbf{Experimental configuration.} The detector fibers runs through the center of a cylindrical block. The sample is put on top, surrounded by a ring of having an inner diameter matching the sample. To reduce stray light, parts are made in black Delrin plastic. Part (a) shows arrangement spatially resolved PTOFS. Centering of sample and ring sets $r_d=0$.  Sideway translation of the source fiber allows tuning of $r_s$. Part (b) shows how transverse translation of sample and ring allow simultaneous tuning of $r_d$ and $r_s$, a configurations used for investigation of sample homogeneity (keeping $\delta r=0$).}\label{fig:setup}
\end{figure}

Light transport is investigated by conducting spatially resolved photon time-of-flight spectroscopy (PTOFS). Short (picosecond) pulses of 760 nm light are injected into the tablets at the radial position $r_s$ using a $600~\mu$m-core graded-index optical fiber, and light transmitted through the sample is detected at the radial positions $r_d$ using a second identical fiber (radial positions defined relative to the sample center). A key parameter in measurements is the radial (transverse) source-detector separation, $\delta r=r_s-r_d$. The collected light consist of pulses that are temporally broadened due to multiple scattering, and the detection fiber direct these pulses to a fast photomultiplier that combined with time-correlated single photon counting allows us to resolve them in time. Details on the spectroscopic instrument is available in Ref. \cite{Svensson2009_RevSciInstrum}, and the experimental configuration is elaborated in Figure \ref{fig:setup}. 

To avoid being very close to the edge of the samples (sample radii being 6.5 mm), measurements were conducted up to $\delta r=5$ mm. For the most strongly scattering samples, however, there was an upper limit to $\delta r$ above which the signal is so weak so that stray light distorts obtained time-of-flight (TOF) histograms. In calcium phosphate samples and the porous alumina, for example, light completely change its direction on average each every $5~\mu$m (i.e., the transport mean free path of light is around $5~\mu$m). The resulting high reflectivity and strong attenuation limits us to measurements up to $\delta r$ of ~3 mm. At larger $\delta r$, measurements are susceptible of stray light, e.g. light that exited the sample at lower $\delta r$ and ideally should not reach the detector fiber. Analysis of the obtained photon TOF histograms are made by employing standard diffusion models generalized for anisotropic diffusion \cite{Wiersma1999_PRL,Wiersma2000_PRE}, and using the extrapolated boundary conditions as described by Contini et al. \cite{Contini1997_ApplOpt}.

\begin{figure}
  \includegraphics[]{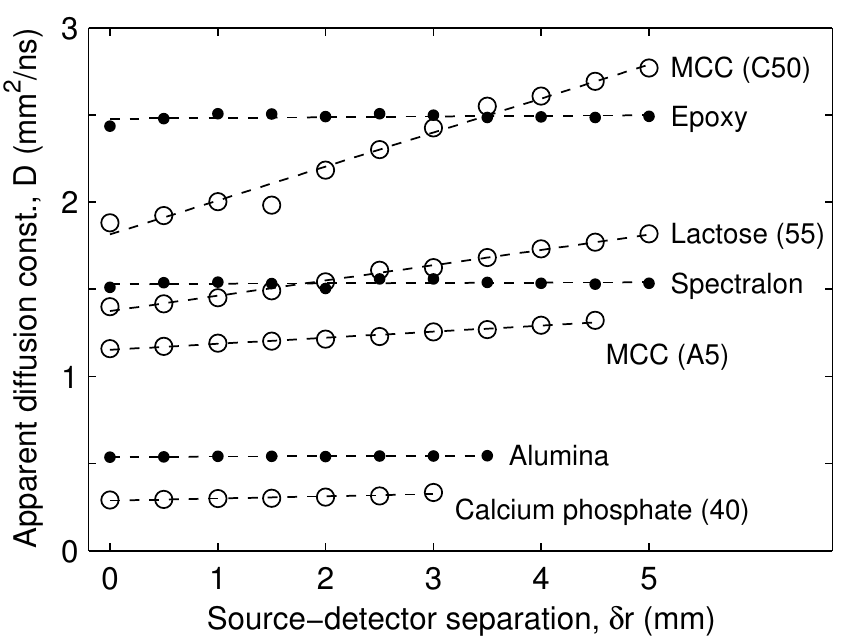}
  \caption{\textbf{The failure of isotropic modeling.} For homogenous isotropic materials, isotropic diffusion modeling should result in the same fitted diffusion constant regardless of the source-detector separation used in the measurement. While this is the case for our isotropic reference materials (solid dots), it is evident that the compacted granular samples (open circles) cannot be modeled in the same manner. For clarity, the graph shows data only from a few of the samples (compression force in kN is given within parantheses).} \label{fig:isotropic_model}
\end{figure}

The inadequacy of homogenous isotropic diffusion models is elucidated in Fig. \ref{fig:isotropic_model}. There, measurements at different source and detector separation are evaluated individually using the isotropic diffusion model. If diffusion is isotropic, fitted diffusion constants should be constant (independent of $\delta r$). Data from the reference materials are thus in excellent agreement with isotropic diffusion, and shows the capability of our system to deliver consistent data over a wide range of optical properties and source detector separations. At the same time, it is evident that the light transport in the compacted granular samples cannot be explained by homogenous isotropic diffusion. Note that also for the calcium phosphate data, although not easily seen in the figure, fitted diffusion constants are non-constant and systematically increasing with $\delta r$ (growing from 0.29 at $\delta r=0$ to 0.33 mm$^2$/ns at $\delta r=3$).

One possible cause of the mismatch with the model of homogenous isotropic diffusion could, of course, be that our samples exhibit a spatially varying diffusion constant due to macroscopic density heterogeneity. However, careful investigation of sample homogeneity rules out this possibility. First, we investigated the radial sample homogeneity optically by making PTOFS measurements with $\delta r=0$ at different positions of the samples ($r_s=r_d$ from 0 to 5 mm in steps of 0.5 mm, cf. Fig. \ref{fig:setup}b). For the compacted granular samples, coefficients of variation in fitted $D$ was smaller than 3\% (similar variation was seen also for the isotropic reference materials), and dependence on radial position was insignificant. This shows that our compacted samples are radially homogenous. To verify longitudinal homogeneity, our structures were also investigated by X-ray microtomography ($\mu$CT). Since radial homogeneity is confirmed by our optical experiments, $\mu$CT scans were optimized for longitudinal density mapping. Still, there were no signs of longitudinal heterogeneity.

\begin{figure}
  \includegraphics[]{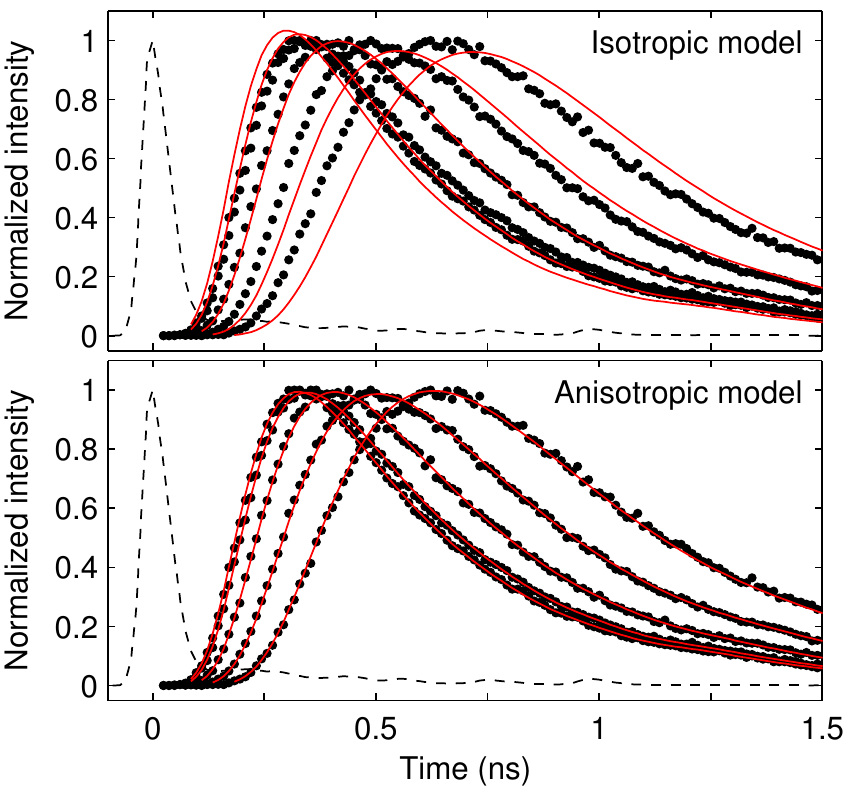}
  \caption{\textbf{The spatio-temporal picture.} When moving from isotropic to anisotropic modeling, fits obtained 
during evaluation of spatially resolved PTOFS (i.e., evaluation of $\delta r$ series) goes from being clear evidence of model inadequacy to becoming virtually perfect. The displayed data is from the MCC C50 sample but is representative for all samples. For visual clarity, only a subset of the involved spatial recordings are shown ($\delta r=0,1,2,3,4$ mm).}\label{fig:simultaneous_fits}
\end{figure}

Convincing evidence that anisotropy is the proper explanation of why homogenous isotropic diffusion modeling fails is presented in Fig. \ref{fig:simultaneous_fits}. There, simultaneous evaluation of the TOF histograms of a $\delta r$ series clearly shows that isotropic homogenous diffusion is far from being capable of explaining experimental data, while the anisotropic diffusion model, with its two diffusion constants, results in excellent fits. Fig. \ref{fig:simultaneous_fits} gives only one example, but similar results were observed for all compressed granular samples. Even for the calcium phosphate samples, which looking at Fig. \ref{fig:isotropic_model} may appear close to isotropic, a dramatic reduction of residuals are obtained when moving to anisotropic diffusion modeling. In comparison, and as expected, anisotropic modeling of data acquired from the isotropic reference materials result in insignificant improvement of fits.

\begin{figure}
  \includegraphics[]{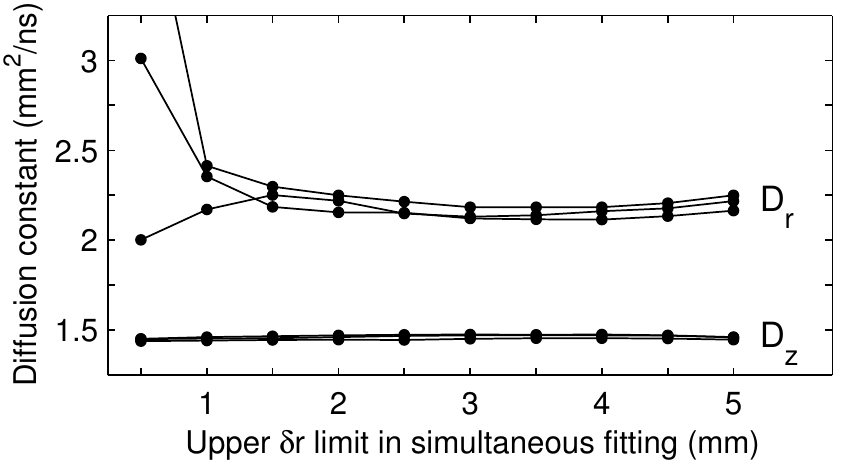}
  \caption{\textbf{The necessity of spatial information.} Typical evolution of fitted radial and longitudinal diffusion coefficients ($D_r$ and $D_z$) as more and more spatial information is included (i.e., evaluation involving all datasets up to an upper $\delta r$). In order to reach good estimates of both longitudinal and radial diffusion constants, evaluation must clearly include measurements made at $\delta r$ being on the order of the sample thickness. Here, datapoints from three measurement repetitions from the MCC A (20kN) sample are shown.}\label{fig:spatial_info}
\end{figure}

It is important to realize that estimation of $D_r$ requires measurements at $\delta r\neq0$, while $D_z$ can be estimated using only a single measurement at $\delta r=0$. This can be inferred directly from anisotropic diffusion theory (cf. Eq. \ref{eq:I_tr}), and the issue is elaborated further in Fig. \ref{fig:spatial_info} by showing how estimates of diffusion constants evolve as the amount of spatial information increase. While estimation of $D_z$ is very robust, proper evaluation of $D_r$ requires inclusion of measurements made at $\delta r$-values comparable to sample thickness. On the other hand, $D_r$ converge to a constant value rather quickly. This means that inclusion of very large $\delta r$-values is not necessary in general, making it possible to avoid low transmission measurements where stray light may distort data. Measurements made very close to the radial boundary may also be influenced by edge effects in material density, and eventually also by the breakdown of the infinite slab diffusion-model. In fact, seen in Fig. \ref{fig:spatial_info}, we generally observe a minor change in $D_r$ when including measurements at high $\delta r$ (when $\delta r=5$ mm, the outer part of the fiber core is only 1.2 mm away from the sample boundary). In the following, our evaluation of anisotropy is therefore based on simultaneous evaluation of TOF histograms recorded at $\delta r$ not larger than 4 mm (as mentioned earlier, the strongly scattering calcium phosphate samples are exceptions, and evaluation is limited to $\delta r$ not larger than 3 mm).

\begin{figure}
  \includegraphics[]{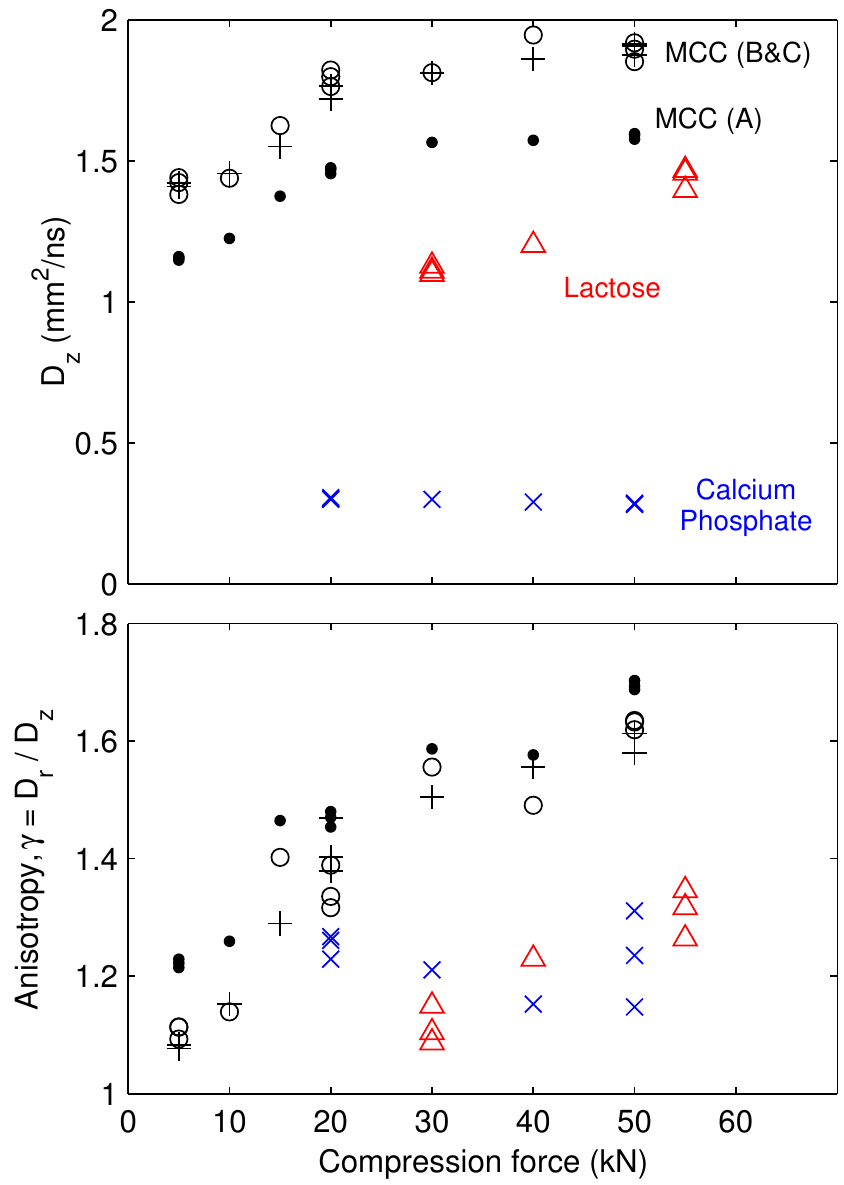}
  \caption{\textbf{Anisotropy summary.} Longitudinal diffusion constants and anisotropy factor $\gamma = D_r/D_z$, versus compression force for all samples. The general trend is that both anisotropy and diffusion constants increase with pressure (the increase in $D$ corresponds to a decrease of the scattering strength). MCC samples are shown as solid circles ($\bullet$), plus signs (+) and open circles ({\large $\circ$}), for the granule size classes A, B and C respectively. Lactose samples are given by triangles ({\footnotesize $\triangle$}), and calcium phosphate samples by the times symbol ($\times$). Note that three measurement repetitions were done on samples made at the lowest and highest compression force, as well as for the intermediate 20 kN MCC samples.}\label{fig:summary}
\end{figure}

Fig. \ref{fig:summary} presents observed anisotropy and longitudinal diffusion constants for all the materials. Measurement repetitions (also shown) show that the variation in derived $D_z$, $D_r$ and $\gamma$ generally is very small. The coefficients of variation were between 0.3 and 3.3\%, except for the 50 kN calcium phosphate which gave a 6.6\% percent variation in $D_r$ and $\gamma$). The anisotropy of the MCC and lactose samples increase with pressure. Interestingly, this trend is not shared by the calcium phosphate samples.  In this context, it is important to note also differences in scattering properties. It is clear that the scattering strength is largely the same for the calcium phosphates. In contrast, scattering decrease with pressure for lactose and MCC samples (note, however, the plateau reached at ~30 kN for MCC samples, after which scattering remains constant).

\section*{DISCUSSION}

The fact that light propagation in compacted granular media can be strongly anisotropic represents a major revision of our understanding of the optics of such systems.  It is therefore important to realize that our findings correlate well with recent reports on anisotropic pore structure. For example, pore space anisotropy of MCC compacts have been confirmed and shown to increase with compression force \cite{Busignies2008_EJPB,Porion2010_PharmRes}. In addition, the pore space anisotropy reported in these studies occur on length scales that should influence light scattering. The fact that radial diffusion is faster than longitudinal is presumably related to formation of crack-like pores running perpendicular to the compression axis \cite{Wu2008_PharmSciTech}.  On the other hand, it should be noted that NMR porosimetry has been applied to e.g. lactose and calcium phosphate samples without clear signs of anisotropy. However, the major difference in the way the the the pore space is probed (and differences in samples) makes deeper comparison meaningless. We note, instead, that anisotropy of \textsl{mechanical} properties have been observed for compacts based on all material investigated here \cite{Ando1983_ChemPharmBull,Mullarney2006_IntJPharm}, rendering our observations highly plausible. In fact, since light scattering and its wavelength dependence is extremely sensitive to microstructure, we believe that spectroscopic optical methods may turn out to be very useful and sensitive probes of pore space anisotropy.  Conventional porosimetry methods like mercury intrusion and gas adsorption does not give information of pore directionality, and currently employed methods that do are subject to important limitations \cite{Andersson2009_PowderTechnol}.

Besides opening a promising new route for characterization of material anisotropy, our results also have obvious implications for quantitative spectroscopy of compacts in general. The use of isotropic diffusion models must be reconsidered. Moreover, considering a combination of anisotropic effects with the density variations characteristic of more complex shapes of e.g. pharmaceutical tablets \cite{Sinka2004_IntJPharm}, it is clear that model-based quantitative spectroscopy of powder compacts indeed is challenging. The picture gets even more complicated when considering also the more fundamental issue of how light sample complex porous structures \cite{Faez2009_PRL,Svensson2010_OptLett}. 

To conclude, our observation of strong anisotropic diffusion of light in compacted granular media have broad implications.  As discussed above, the phenomenon can be used for material characterization and fundamental investigations of compression physics, and also has important implications for quantitative spectroscopy of compacts. At the same time, reaching full understanding of the relation between compression, microstructure and light scattering is challenging, and this endeavor will require extensive and interdisciplinary efforts in the future.

\section*{METHODS}

\subsection*{Compacted granular samples}
The compacted granular samples are based on either microcrystalline cellulose (MCC), lactose or calcium phosphate. All compacts were manufactured using a standard single-punch tablet press with a cylindrical die (all having a 13 mm diameter and a total weight of 500 mg). The MCC-based samples represent realistic pharmaceutical tablets, containing 38.1\% MCC (excipient), 31.8\% metoprolol (active ingredient), 12.7\% sodium starch glycolate (disintegrant), 11.1\% lactose (filler), 2.9\% colloidal silica, 2.2\% povidone (binder) 1.2\% magnesium stearate (lubricant, reduces friction during compaction). In order to investigate influence of granule sizes, MCC compacts were manufactured from sieved granules (granule sizes: (A) $<150~\mu$m, (B) $150-400~\mu$m, and (C) $>400~\mu$m, respectively). Compression forces investigated are 5, 10, 15, 20, 30, 40 and 50 kN, yielding sample thicknesses from 3.5 mm down to 2.7 mm. The samples are similar to those in Ref. \cite{Svensson2010_OptLett}, for which mercury intrusion porosimetry showed than predominate pore size is around 1-5 $\upmu$m. Lactose and calcium phosphate samples, on the other hand, were manufactured from pure powders with a 5-10 $\mu$m mean particle size. Compression forces were 30, 40 and 55 kN for lactose (thickness range 2.7-2.6 mm), and 20, 30, 40 and 50 kN for calcium phosphate (thickness range 2.1-1.9 mm).

\subsection*{Diffusion modeling}
In a system with cylindrical symmetry, the time evolution of the intensity transmitted through a slab at a radial distance $\delta r$ from the incident light pulse, $I_\textrm{T}(t,\delta r)$, is given by Eq. \ref{eq:I_tr}  \cite{Wiersma1999_PRL,Wiersma2000_PRE}.

There $I_0$ is the incident intensity, $c$ the speed of light in the medium, $\mu_a$ the absorption coefficient, and $D_r$ and $D_z$ the radial and longitudinal diffusion constants respectively. The quantities $z_{+,m}$ and $z_{-,m}$ contain the dependence on slab thickness $L$. More specifically, $z_{+,m}=2m(L+2z_e)+z_0$ and $z_{-,m}=2m(L+2z_e)-2z_e+z_0$. Apart from $z_{+,0}=z_0$, which gives the position of the isotropic light source that models injected light, they correspond to positions of negative and positive virtual sources introduced to handle boundary effects \cite{Contini1997_ApplOpt}. The parameter $z_e$ is the position of the, so called, extrapolated boundary. It is important to keep in mind that both $z_0$ and $z_e$ depend on the diffusion constant $D_z$. The source is located one transport mean free path into the medium, i.e. $z_0=3D_z/c$, and $z_e$ depends on the effective refractive index ($n_e$) in a more complex manner. In fact, $n_e$ is not known exactly for our complex porous systems, but as long as sample dimension significantly exceed the transport mean free path of light, the time evolution of the transmitted intensity depends only weakly on $z_0$ and $z_e$. We use the approximate value $n_e=1.5$ throughout this work, resulting in a $z_e$ of $3.6279 \times 2D_z/c$. Note that this value of $z_e$ is calculated using ordinary isotropic considerations  \cite{Contini1997_ApplOpt}, and is only a convenient approximation for anisotropic media. We have verified that a change in the assumed value of $n_e$ has negligible influence of results presented. It should also be noted that the size of the fiber cores ($600~\mu$m), i.e. the fact that $\delta r$ is not really a single value for a single measurement, are taken into account by integration over source and detector areas. Finally, note that the isotropic model of dynamic light diffusion through a slab is reached simply by setting $D_z=D_r=D$, where $D$ then is the isotropic diffusion constant of the medium. Fitting of Eq. \ref{eq:I_tr} to experimental data is made using only diffusion constant(s), absorption coefficient and amplitude coefficients as free fit parameters (since $z_0$ and $z_e$ depend on $D_z$, they are varied accordingly). When multiple TOF histograms are evaluated simultaneously (modeling of $\delta r$ series as one compound dataset), each single dataset has its own free amplitude coefficient. Measuring diffuse intensity in an absolute manner is extremely difficult, and attempts to maintain relative intensity relation for different $\delta r$ would be highly inefficient due to the enormous difference in transmission. The absolute time scale ($t=0$) is inferred from careful measurements of the instrumental response function (IRF), and the shape of the IRF is taken into account during fitting by measuring the fit in terms of agreement between experimental data and the convolution between IRF and diffusion model \cite{Svensson2009_RevSciInstrum}.

\begin{widetext}
\begin{align}\label{eq:I_tr}
&I_\textrm{T}(t,\delta r)= \frac{I_0 \exp\left(-\frac{\delta r^2}{4D_{r}t}-\mu_a c t\right)}{\pi^{3/2}(4t)^{5/2}D_{r}\sqrt{D_{z}}} \times \sum_{m=-\infty}^{+\infty} z_{+,m} \exp\left(\frac{-z_{+,m}^2}{4D_{z}t}\right) -z_{-,m} \exp\left(\frac{-z_{-,m}^2}{4D_{z}t}\right)
\end{align}
\end{widetext}

\section*{ACKNOWLEDGEMENTS} 
This work was supported and generously funded by Stefan Andersson-Engels using grants from the Swedish Research Council. Jonas Johansson is acknowledged for assisting in sample manufacturing. Charles Mark Bee and the Imaging Technology Group at the Illinois Beckman Institute are acknowledged for conducting the X-ray microtomography. We are also grateful to Diederik Wiersma for encouragement and for reading the manuscript.

\end{document}